\documentclass[prd, 10pt, reprint, showpacs, showkeys] {revtex4-1}
\usepackage{amsmath}
\usepackage{relsize}
\usepackage{amssymb}
\usepackage{graphics}
\usepackage{epsfig}
\usepackage{epstopdf}
\usepackage{verbatim}
\usepackage{color}
\usepackage{multirow}
\usepackage{subfigure}
\usepackage[toc,page]{appendix}
\begin{document}
\title{Connecting Majorana phases to the Geometric Parameters of Majorana Unitarity Triangle in a model of Neutrino Mass Matrix }
\author{Surender Verma}
\email{sverma@cuhimachal.ac.in, s\_7verma@yahoo.co.in}
\author{Shankita Bhardwaj}
\email{shankita.bhardwaj982@gmail.com}
\affiliation{Department of Physics and Astronomical Science, Central University of Himachal Pradesh, Dharamshala, India 176215}
\date{\today}
\begin{abstract}
We have investigated a possible connection between the Majorana phases and geometric parameters of leptonic unitarity triangle(LUT) in two-texture zero neutrino mass matrix. Such analytical relations can, also, be obtained for other theoretical models viz. hybrid textures, neutrino mass matrix with vanishing minors and have profound implications for geometric description of $CP$ violation. As an example, we have considered two-texture zero neutrino mass model to obtain relation between Majorana phases and LUT parameters. In particular, we find that Majorana phases depend on only one of the three interior angles of LUT in each class of two-texture zero neutrino mass matrix. We have, also, constructed LUT for class $A$, $B$ and $C$ neutrino mass matrices. Non-vanishing areas and nontrivial orientations of these Majorana unitarity triangles indicate non-zero $CP$ violation as a generic feature of this class of mass models.
\end{abstract}
\maketitle
\section{Introduction}
Recent developments in experimental neutrino physics have catalyzed assiduous efforts to understand the origin of neutrino mass. The neutrino oscillation experiments have, now, revealed the dominant structure of the neutrino mixing matrix\cite{acc, reactor, SNO, SK} and oscillation parameters such as, 
$\theta_{12}, \theta_{23}, \theta_{13}, \Delta m_{21}^2, |\Delta m_{31}^2|$ are known to an unprecedented  accuracy. Beside these astonishing discoveries, there remains arduous questions viz. Are neutrinos Majorana or Dirac-particle? What is the lightest neutrino mass? Is neutrino mass hierarchy normal or inverted? What is the octant of atmospheric mixing angle? and what is the leptonic $CP$ violating phase $\delta$?, to name a few. The primary goal of the future neutrino experimental endeavours will be to address these questions by employing a miscellany of experimental configurations and techniques.  

The observation of non-zero value of reactor mixing angle $\theta_{13}$\cite{theta13}, by oscillation experiments, provides unique opportunity for the possible measurement of $CP$ violation in the leptonic sector. The flavor oscillations implies the existence of mixing in the weak charged current interaction
\begin{equation}
\mathcal{L}_{CC}=-\frac{g}{\sqrt{2}}\sum_{l=e,\mu,\tau} \bar{l_L}(x)\gamma_\alpha\nu_{lL}(x)W^{\alpha\dagger}(x)+h.c.,
 \end{equation}
\begin{equation}
\nu_{lL}(x)=\sum_{j=1}^{n} U_{lj}\nu_{jL}(x), \nonumber
\end{equation}
where, $\nu_{lL}(x)$ are flavor fields, $\nu_{jL}(x)$ are $LH$ component of mass fields $\nu_j$ with mass $m_j$, and $U$ is unitary mixing matrix. The mixing matrix $U$ can be parameterised in terms of $\frac{1}{2}n(n-1)$ angles and $\frac{1}{2}(n-1)(n-2)$ phases. For Majorana neutrinos, there exist $(n-1)$ additional $CP$ violating phases called ``Majorana phases". In the standard framework of three-neutrino the mixing matrix contains two $CP$ violating phases $\rho$ and $\sigma$, in addition to one Dirac-type $CP$ violating phase $\delta$. While the Dirac-type $CP$ violating phase will, possibly, be measured in the neutrino oscillation experiments T2K, NO$\nu$A and DUNE, the information about Majorana-type $CP$ violating phases can be extracted from the lepton number violating processes.

$CP$ violation in the leptonic sector can either be studied through the construction of Leptonic Unitarity Triangle(LUT) or by direct measurement of the $CP$ violating phase $\delta$ in the neutrino oscillation experiments. The first approach has the advantage of being rephasing invariant  description of the $CP$ violation. 

In the flavor basis, the unitary mixing matrix $V\equiv UP$, where $U$ is Pontecorvo-Maki-Nakagawa-Sakata(PMNS) matrix, is given by
\begin{equation}
U=\left(
\begin{array}{ccc}
c_{12}c_{13} & s_{12}c_{13} & s_{13}e^{-i\delta} \\~\label{v}
-s_{12}c_{23}-c_{12}s_{23}s_{13}e^{i\delta} &  
c_{12}c_{23}-s_{12}s_{23}s_{13}e^{i\delta} & s_{23}c_{13} \\
s_{12}s_{23}-c_{12}c_{23}s_{13}e^{i\delta} &
-c_{12}s_{23}-s_{12}c_{23}s_{13}e^{i\delta} & c_{23}c_{13}
\end{array}
\right).
\end{equation}
$s_{ij}=\sin\theta_{ij}$, $c_{ij}=\cos\theta_{ij}$ and $P=diag(1,e^{i \rho},e^{i(\sigma+\delta)})$. The unitarity of $V$ imposes six orthogonality conditions on the elements of $V$ viz.,
\begin{eqnarray}
\nonumber
&&\Delta_{e\mu}\equiv V_{e1}V_{\mu1}^{*}+V_{e2}V_{\mu2}^{*}+V_{e3}V_{\mu3}^{*}=0,\\ ~\label{dt}
&&\Delta_{\mu\tau}\equiv V_{\mu1}V_{\tau1}^{*}+V_{\mu2}V_{\tau2}^{*}+V_{\mu3}V_{\tau3}^{*}=0,\\ \nonumber
&&\Delta_{\tau e}\equiv V_{\tau1}V_{e1}^{*}+V_{\tau2}V_{e2}^{*}+V_{\tau3}V_{e3}^{*}=0,\\ \nonumber
\end{eqnarray}
obtained from multiplying two rows of $V$ and
\begin{eqnarray}
\nonumber
\Delta_{12}\equiv V_{e1}V_{e2}^{*}+V_{\mu1}V_{\mu2}^{*}+V_{\tau1}V_{\tau2}^{*}=0,\\ ~\label{mt}
\Delta_{23}\equiv V_{e2}V_{e3}^{*}+V_{\mu2}V_{\mu3}^{*}+V_{\tau2}V_{\tau3}^{*}=0,\\ \nonumber
\Delta_{31}\equiv V_{e3}V_{e1}^{*}+V_{\mu3}V_{\mu1}^{*}+V_{\tau3}V_{\tau1}^{*}=0,\\ \nonumber
\end{eqnarray}
obtained from multiplying two columns of $V$. In the complex plane, Eq.(\ref{dt}) describes the ``Dirac triangles(DT)" whereas Eq.(\ref{mt}) describes ``Majorana triangles(MT)". Under the rephasing transformations of the leptonic fields, the mixing matrix $V$ transforms as $V_{li}\rightarrow e^{i\phi_i}V_{li}$. Thus, the Dirac unitarity conditions, Eq.(\ref{dt}), rotate in the complex plane and their orientations have no physical significance. In general, vanishing area of DT does not, necessarily, imply vanishing $CP$ violation because of the non-zero contribution coming from the Majorana sector. The orientations of Majorana unitarity triangles are physical because, as can be seen from Eq.(\ref{mt}), they corresponds to sum of bilinear-rephasing invariant terms. Thus, Majorana triangles provide a complete description of the $CP$ violation. 

Although the dominant structure of the PMNS matrix has been revealed but $CP$ violating phases are still unknown. There are various theoretical constructions to restrict the number of free parameters in the leptonic flavor sector to make a particular model of neutrino masses and mixings more predictive. These approaches include texture zeros\cite{tz1,tz2,tz3}, hybrid texture\cite{ht1,ht2} and vanishing minor\cite{vm1,vm2} which are consistent with current data on neutrino masses and mixings. These ansatze have profound phenomenological implications for our quest to understand the origin of neutrino masses and mixings. In the present work, we have considered two-texture zero model of neutrino mass matrix  and derive possible connection between the Majorana phases and geometric parameters of Majorana unitarity triangle. 

In Sec. II, we have reviewed the two-texture zero mass model and obtained mass ratios to study the complete phenomenology of the model. We derive relation between Majorana phases and geometric parameters of Majorana unitarity triangles, and construct these triangles for all viable two-texture zero neutrino mass matrices in Sec. III. In Sec. IV, we have investigated the status of $0\nu\beta\beta$ decay in two-texture zero neutrino mass models. In Sec. V we have discussed the generic features of the MTs and, finally, we conclude in Sec. VI.

\section{Two-Texture Zeros Neutrino Mass Matrices}
In the flavor basis, where the charged-lepton mass matrix $M_{\nu}$ is diagonal, the Majorana neutrino mass matrix $M$ is given by
\begin{equation}\label{matrixM}
M=V M_{\nu} V^{T},
\end{equation}
where, $M_{\nu}=Diag\lbrace m_{1},m_{2},m_{3}\rbrace$. There are total 15 possible patterns of two-texture zero Majorana neutrino mass matrices. Seven out of these patterns, are found to be consistent with the neutrino mixing parameters. These are $A_{1},A_{2},B_{1},B_{2},B_{3},B_{4}$ and $C$(Table \ref{tab1}). Since $M$ is symmetric it has total six independent complex elements. Any two vanishing elements of $M$, i.e., $ M_{st}=0,M_{xy}=0,$
where $s, t, x,$ and $y$ can take values $e,\mu,\tau$, result in two complex constraining equations viz;
$\sum_{i=1}^{3}V_{si}V_{ti}m_{i}=0, \hspace{3mm} \sum_{i=1}^{3}V_{xi}V_{yi}m_{i}=0$. 
These two relations involve nine free parameters i.e., $m_{1},m_{2},m_{3}$, $\theta_{12},\theta_{13},\theta_{23}$ and three $CP$-violating phases $\rho,\sigma$ and $\delta$, where $\rho, \sigma$ are two Majorana-type $CP$ violating phases and $\delta$ is Dirac-type $CP$ violating phase. We solve these constraining equations to obtain mass ratios($\frac{m_{1}}{m_{2}},\frac{m_{1}}{m_{3}}$) and Majorana phases($\rho,\sigma$) as,

  \begin{eqnarray}\label{m12}
  \frac{m_{1}}{m_{2}}&=&\left|\frac{U_{x2}U_{y2}U_{s3}U_{t3}-U_{s2}U_{t2}U_{x3}U_{y3}}{U_{s1}U_{t1}U_{x3}U_{y3}-U_{s3}U_{t3}U_{x1}U_{y1}}\right|,
 \end{eqnarray}

\begin{eqnarray}\label{m13} 
\frac{m_{1}}{m_{3}}&=&\left|\frac{U_{x3}U_{y3}U_{s2}U_{t2}-U_{s3}U_{t3}U_{x2}U_{y2}}{U_{s1}U_{t1}U_{x2}U_{y2}-U_{s2}U_{t2}U_{x1}U_{y1}}\right|,
\end{eqnarray}
and
\begin{eqnarray}\label{rho}
\rho=-\frac{1}{2}Arg\left(\frac{U_{x2}U_{y2}U_{s3}U_{t3}-U_{s2}U_{t2}U_{x3}U_{y3}}{U_{s1}U_{t1}U_{x3}U_{y3}-U_{s3}U_{t3}U_{x1}U_{y1}}\right),
\end{eqnarray}

\begin{eqnarray}\label{sigma}
\sigma=-\frac{1}{2}Arg\left(\frac{U_{x3}U_{y3}U_{s2}U_{t2}-U_{s3}U_{t3}U_{x2}U_{y2}}{U_{s1}U_{t1}U_{x2}U_{y2}-U_{s2}U_{t2}U_{x1}U_{y1}}\right)-\delta,
\end{eqnarray}
respectively.
\begin{table*}
\caption{\label{tab1}The mass ratios for each type of neutrino mass matrices upto first order in $s_{13}$.}
\begin{ruledtabular}
\centering
\begin{tabular}{ p{0.5 \linewidth} p{0.5 \linewidth}}
Type  &  \hspace{1cm} Mass Ratios\\
of Texture&\\
\hline
 $ A_{1}(M_{ee}=0;M_{e\mu}=0)$ & $\begin{array}{lcl}\frac{m_{1}}{m_{2}}&=&\tan^{2}\theta_{12}\left(1-\frac{\cot\theta_{23}}{s_{12}c_{12}}s_{13}\cos\delta\right)\\ \frac{m_{1}}{m_{3}}&=&\tan\theta_{12}\tan\theta_{23}s_{13}\end{array}$\\
\hline
 $ A_{2}(M_{ee}=0;M_{e\tau}=0)$ & $\begin{array}{lcl}\frac{m_{1}}{m_{2}}&=& \tan^2\theta_{12}\left(1+\frac{\tan\theta_{23}}{c_{12}s_{12}}s_{13}\cos\delta\right)\\ \frac{m_{1}}{m_{3}}&=&\tan\theta_{12}\cot\theta_{23}s_{13}\end{array}$\\
  \hline
 $ B_{1}(M_{e\tau}=0;M_{\mu\mu}=0)$ & $\begin{array}{lcl}\frac{m_{1}}{m_{2}}&=& 1+\frac{c_{23}}
 {c_{12}s_{12}s_{23}^3}s_{13}\cos\delta\\
 \frac{m_{1}}{m_{3}}&=&\tan^2\theta_{23}\left(1+\frac{\tan\theta_{23}\cot\theta_{12}}{s_{23}^{2}}s_{13}\cos\delta\right)\end{array}$\\
   \hline
 $B_{2}(M_{e\mu}=0;M_{\tau\tau}=0)$ & $\begin{array}{lcl}\frac{m_{1}}{m_{2}}&=& 1-\frac{s_{23}}
 {c_{12}s_{12}c_{23}^3}s_{13}\cos\delta\\
 \frac{m_{1}}{m_{3}}&=&\frac{1}{\tan^{2}\theta_{23}}\left(1-\frac{\tan\theta_{23}\cot\theta_{12}}{c_{23}^2}s_{13}\cos\delta\right)\end{array}$\\
   \hline
 $B_{3}(M_{e\mu}=0;M_{\mu\mu}=0)$  & $\begin{array}{lcl}\frac{m_{1}}{m_{2}}&=& 1-\frac{c_{23}\tan^2\theta_{23}}{c_{12}s_{12}s_{23}^{3}}s_{13}\cos\delta\\ \frac{m_{1}}{m_{3}}&=&\tan^{2}\theta_{23}\left(1-\frac{\cot\theta_{12}}{s_{23}c_{23}}s_{13}\cos\delta\right) \end{array}$\\
   \hline
 $B_{4}(M_{e\tau}=0;M_{\tau\tau}=0)$  & $\begin{array}{lcl}\frac{m_{1}}{m_{2}}&=& 1+\frac{s_{23}\cot^{2}\theta_{23}}{c_{12}s_{12}c_{23}^{3}}s_{13}\cos\delta\\
     \frac{m_{1}}{m_{3}}&=&\frac{1}{\tan^{2}\theta_{23}}\left(1+\frac{\cot\theta_{12}}{s_{23}c_{23}}s_{13}\cos\delta\right)\end{array}$\\
   \hline
 $C(M_{\mu\mu}=0;M_{\tau\tau}=0)$      &  $\begin{array}{lcl}\frac{m_{1}}{m_{2}}&=&
 \frac{1}{\tan^{2}\theta_{12}}\left(1-\frac{\tan\theta_{23}}{s_{12}c_{12}}s_{13}\cos\delta\right)\\
      \frac{m_{1}}{m_{3}}&=&\frac{1}{\tan\theta_{12}\tan2\theta_{23}s_{13}} 
      \left(1+\frac{4(-s_{12}^{2}+c_{12}^{2}\cos^{2}2\theta_{23})}{\sin4\theta_{23}\sin\theta_{12}}s_{13}\cos\delta\right)\end{array}$\\
\end{tabular}
\end{ruledtabular}
\end{table*}
 Using Eqs.(\ref{v}), (\ref{m12}) and (\ref{m13}), we have shown these mass ratios upto first order in $s_{13}$ in Table \ref{tab1}. These relations will be  useful while studying the phenomenology of two-texture zero neutrino mass matrices. In our numerical analysis, we have used global data\cite{2016data} to obtain the allowed parameter space of neutrino mixing parameters in two-texture zero neutrino mass model. The best fit point(bfp) and $1\sigma$ range of these parameters are tabulated in Table \ref{tab2}. We have, also, calculated the Jarlskog\cite{jarlskog} $CP$ invariant $J_{CP}$ sensitive to Dirac phase $\delta$
 \begin{equation}
 J_{CP}=Im\{V_{e1}V_{\mu2}V_{e2}^{*}V_{\mu1}^{*}\},
 \end{equation}
 and $s_1$, $s_2$ $CP$ invariants\cite{s1s21, s1s22}  
 \begin{eqnarray}
s_{1}&&=Im\{V_{e1}V_{e3}^{*}\},\\
s_{2}&&=Im\{V_{e2}V_{e3}^{*}\},
 \end{eqnarray}
 sensitive to Majorana phases $\rho$ and $\sigma$ (Table \ref{tab3}). 
 
\begin{table*}
 \caption{\label{tab2} The neutrino mixing parameters in two-texture zero neutrino mass matrices.}
\begin{ruledtabular}
  \centering
 \begin{tabular}{c  l  l}
   \multirow{2}{*}{Type of Texture} & \multicolumn{2}{c}{\textbf{bfp $\pm1\sigma$   in degrees($^{o}$)}} \\
    \cline{2-3}
     & \multicolumn{1}{c}{Normal Hierarchy} &  \multicolumn{1}{c}{Inverted Hierarchy}   \\
    \hline
    $A_{1}$ & $\theta_{12}=33.52^{+0.74}_{-0.74},\theta_{13}=8.45_{-0.14}^{+0.14},\theta_{23}=41.63^{+1.38}_{-1.38},$& \multicolumn{1}{c} -\\
        &$ \rho=-77.42^{+34.71}_{-34.71},\sigma=-102.60^{+48.37}_{-48.37},\delta=67.58^{+31.99}_{-31.99}. $ &  \\ 
             \hline
   $ A_{2}$ & $\theta_{12}=34.55^{+0.57}_{-0.57},\theta_{13}=8.38^{+0.14}_{-0.14},\theta_{23}=43.27^{+1.07}_{-1.07}$,& \multicolumn{1}{c} -\\
                  &$\rho=86.69^{+1.91}_{-1.91},\sigma=-78.44^{+6.68}_{-6.68},\delta=157.1^{+15.77}_{-15.77}. $ &    \\ 
    \hline
    $B_{1}$ & $\theta_{12}=33.75^{+0.57}_{-0.57},\theta_{13}=8.45^{+0.14}_{-0.14},\theta_{23}=41.10^{+1.07}_{-1.07}$,&$\theta_{12}=33.52^{+0.77}_{-0.77},\theta_{13}=8.47^{+0.13}_{-0.13},\theta_{23}=45.76^{+0.52}_{-0.52},$\\
                  &$\rho=-3.67^{+1.73}_{-1.73},\sigma=-0.36^{+0.25}_{-0.25},\delta=267.10^{+15.21}_{-15.21}. $ & $\rho=0.47^{+0.32}_{-0.32},\sigma=-177.36^{+28.84}_{-28.84},\delta=268.87^{+14.47}_{-14.47}. $\\
                      \hline 
    $B_{2}$ & $\theta_{12}=33.60^{+0.76}_{-0.76},\theta_{13}=8.45^{+0.14}_{-0.14},\theta_{23}=45.75^{+0.47}_{-0.47}$,&$\theta_{12}=33.71^{+0.76}_{-0.76},\theta_{13}=8.45^{+0.14}_{-0.14},\theta_{23}=41.03^{+1.26}_{-1.26},$\\
                 &$ \rho=0.46^{+0.50}_{-0.50},\sigma=-174.92^{+42.41}_{-42.41},\delta=266.91^{+21.29}_{-21.29}. $ &$\rho=-1.85^{+0.55}_{-0.55},\sigma=-1.45^{+0.45}_{-0.45},\delta=269.00^{+13.24}_{-13.24} .$ \\
                      \hline
   $ B_{3}$ &$\theta_{12}=33.64^{+0.74}_{-0.74},\theta_{13}=8.49^{+0.14}_{-0.14},\theta_{23}=41.10^{+1.30}_{-1.30},$&$\theta_{12}=33.58^{+0.75}_{-0.75},\theta_{13}=8.47^{+0.14}_{-0.14},\theta_{23}=45.78^{+0.52}_{-0.52},$\\
                 &$ \rho=2.43^{+0.97}_{-0.97},\sigma=-177.50^{+31.89}_{-31.89},\delta=267.80^{+16.31}_{-16.31}. $  &$\rho=-0.47^{+0.32}_{-0.32},\sigma=-0.31^{+0.18}_{-0.18},\delta=269.47^{+9.20}_{-9.20}. $ \\ 
    \hline
   $ B_{4}$ & $\theta_{12}=33.66^{+0.71}_{-0.71},\theta_{13}=8.45^{+0.15}_{-0.15},\theta_{23}=45.87^{+0.54}_{-0.54},$ & $\theta_{12}=33.74^{+0.74}_{-0.74},\theta_{13}=8.45^{+0.14}_{-0.14},\theta_{23}=41.29^{+1.27}_{-1.27},$\\
                 &$ \rho=-0.47^{+0.34}_{-0.34},\sigma=-0.25^{+0.77}_{-0.77},\delta=268.50^{+10.90}_{-10.90}. $ &
                 $ \rho=2.24^{+0.90}_{-0.90},\sigma=-167.4^{+30.86}_{-30.86},\delta=266.82^{+15.60}_{-15.60}. $    \\ 
    \hline
   $ C$ & \multicolumn{1}{c} - & $\theta_{12}=33.54^{+0.75}_{-0.75},\theta_{13}=8.44^{+0.14}_{-0.14},\theta_{23}=41.08^{+1.26}_{-1.26}$,\\
                  & &$\rho=46.65^{+12.33}_{-12.33},\sigma=-165.06^{+19.07}_{-19.07},\delta=292.80^{+15.17}_{-15.17}. $ \\
  \end{tabular}
     \end{ruledtabular}
\end{table*}

\begin{table*}
	\caption{\label{tab3}The $CP$ invariants in two-texture zero neutrino mass matrices.}
	\begin{ruledtabular}
		\centering
		\begin{tabular}{ c c c c c}
			Type of Texture & Hierarchy &$ J_{CP}$(bfp$\pm1\sigma)$& $s_{1}$(bfp$\pm1\sigma)$ & $s_{2}$(bfp$\pm1\sigma)$ \\
			\hline
			$ A_{1}$ & NH & $(2.67\pm1.16)\times10^{-2}$  & $(-0.10\pm 0.04)\times10^{-2}$ & $(-0.14\pm0.06)\times10^{-2}$\\
			\hline
			$ A_{2}$ & NH & $(1.60\pm0.80)\times10^{-2}$  & $(-0.28\pm0.03)\times10^{-2}$ & $(0.40\pm0.03)\times10^{-2}$\\
			\hline
			$ B_{1}$ & NH & $(3.21\pm0.63)\times10^{-2}$& $(-1.38\pm0.98)\times10^{-5}$ & $(-8.30\pm3.60)\times10^{-5}$      \\
			
			         & IH & $(3.22\pm0.85)\times10^{-2}$ & $(0.64\pm0.17)\times10^{-2}$&  $(0.43\pm0.11)\times10^{-2}$    \\
			\hline
			$B_{2}$  & NH & $(3.29\pm0.34)\times10^{-2} $ & $(-0.66\pm0.06)\times10^{-2} $& $(0.44\pm0.04)\times10^{-2}$    \\
			
			         & IH & $(3.22\pm0.62)\times10^{-2} $ & $ (-4.97\pm1.79)\times10^{-5} $ & $(-1.43\pm0.44)\times10^{-5}$     \\
			\hline
			$B_{3}$  &  NH    & $(3.22\pm0.58)\times10^{-2} $ & $(-6.54\pm1.19)\times10^{-3}$ & $(4.41\pm0.80)\times10^{-3}$     \\
			
			          & IH  & $(3.29\pm0.68)\times10^{-2}$  &$(-1.65\pm0.65)\times10^{-4}$& $(-1.01\pm0.65)\times10^{-4}$    \\
			\hline
			 $B_{4}$&  NH   &$(3.32\pm0.53)\times10^{-2}$   & $(-1.99\pm0.94)\times10^{-6}$ & $(-1.20\pm0.87)\times10^{-5}$ \\
			 
			        &  IH   & $(3.23\pm0.55)\times10^{-2}$    &  $(-6.50\pm1.10)\times10^{-3}$ & $(4.30\pm0.75)\times10^{-3}$\\
			   \hline
			   $C$& IH   &  $(2.95\pm0.38)\times10^{-2}$   & $(-6.19\pm0.64)\times10^{-3}$  &  $(5.25\pm0.59)\times10^{-3}$   \\
		\end{tabular}
	\end{ruledtabular}
\end{table*}

\section{Majorana Unitarity Triangles in Two-Texture Zero Model}
In Majorana triangles(Eq.(\ref{mt})), all the terms are rephasing invariant and hence do not rotate in complex plane under rephasing transformations. The study of these triangles, is of great physical significance due to their dependence on Majorana phases\cite{ut, xing}.  The MTs $\Delta_{12}$ and $\Delta_{31}$ are sensitive to $\rho$ and $\sigma$, respectively, whereas, $\Delta_{23}$ is sensitive to both Majorana phases ($\rho,\sigma$). In general, the sides and angles of MT can be expressed as,
\begin{equation}\label{sides}
 \left(S_{1},S_{2},S_{3}\right)=\left(|V_{ef}V_{ef'}|,|V_{\mu f}V_{\mu f'}|,|V_{\tau f}V_{\tau f'}|\right),
\end{equation}
\vspace{1mm}
\begin{eqnarray}\label{angles}
\nonumber
\alpha_{ff'}&&=Arg\left(-\frac{V_{\mu f}V_{\mu f'}^{*}}{V_{\tau f}V_{\tau f'}^{*}}\right),
 \beta_{ff'}=Arg\left(-\frac{V_{\tau f}V_{\tau f'}^{*}}{V_{ef}V_{ef'}^{*}}\right),\\
\gamma_{ff'}&&=Arg\left(-\frac{V_{ef}V_{ef'}^{*}}{V_{\mu f}V_{\mu f'}^{*}}\right),
\end{eqnarray}
where, $S_{1},S_{2},S_{3}$ are three sides and $\alpha,\beta,\gamma$ are three angles with subscript $f, f'=(1,2,3)$ and $f\neq f'$. We have obtained relation between Majorana phases and interior angles of MT for class $A,B$ and $C-$type neutrino mass matrices which are tabulated in Table \ref{tab4}. The procedure to obtain these expressions has been elaborated in the Appendix. As can be seen from the Table \ref{tab4}, the Majorana phases depend only on one of the interior angle of Majorana triangles in two-texture zero neutrino mass model. These relations useful to identify the corresponding Majorana triangle for each type of two-texture zero neutrino mass matrices. Hence, we choose the Majorana triangle $\Delta_{12}$ for $A_{1}$ and $A_{2}$, $\Delta_{23}$ for $B_{1}$ and $B_{2}$, $\Delta_{31}$ for $B_{3},B_{4}$ and $C$.

\begin{table*}[hbtp]
\caption{\label{tab4}The relation between Majorana phases and angles of Majorana unitarity triangle.}
\begin{ruledtabular}
\begin{tabular}{ c p{0.6\linewidth}} 

 Type of Texture & \hspace{0cm}Majorana Phases in terms of angles of Majorana unitarity triangle\\ 
\hline
 $A_{1}$ & $\rho=-\frac{1}{2}\left(\gamma_{12}-\pi+Arg\left(\frac{U_{e3}U_{\mu2}U_{\mu1}U_{e2}^{2}-U_{e2}^{3}U_{\mu3}U_{\mu1}}{U_{\mu2}U_{e1}^{3}U_{\mu3}-U_{\mu2}U_{e3}U_{\mu1}U_{e1}^{2}}\right)\right)$,\\
 &\\ 
    &  $\sigma=-\frac{1}{2}\left(\gamma_{12}-\pi+Arg\left(\frac{U_{e2}^{3}U_{e3}U_{\mu3}U_{\mu1}-U_{e3}^{2}U_{e2}^{2}U_{\mu2}U_{\mu1}}{U_{e1}^{3}U_{e2}U_{\mu2}^{2}-U_{e2}^{2}U_{e1}^{2}U_{\mu1}U_{\mu2}}\right)\right)-\delta$. \\ 
   & \\
\hline
  $A_{2}$ &$\rho=\frac{1}{2}\left(\beta_{12}-\pi+Arg\left(\frac{U_{e3}U_{\tau2}^{2}U_{\tau1}-U_{\tau3}U_{e2}U_{\tau1}U_{\tau2}}{U_{e1}U_{\tau1}U_{\tau2}U_{\tau3}-U_{\tau1}^{2}U_{e3}U_{\tau2}}\right)\right)$,\\
   & \\
    &$\sigma=-\frac{1}{2}\left(\beta_{12}-\pi+Arg\left(\frac{U_{e2}^{3}U_{e3}U_{\tau3}-U_{e3}^{2}U_{e2}^{2}U_{\tau2}}{U_{e1}U_{e2}U_{\tau2}^{2}U_{\tau1}-U_{e2}^{2}U_{\tau1}^{2}U_{\tau2}}\right)\right)-\delta$.\\
    &\\
\hline
  $ B_{1}$ & $\rho=-\frac{1}{2}\left(\beta_{23}-\pi+Arg\left(\frac{U_{\mu2}^{2}U_{e3}U_{\tau3}^{2}U_{e2}-U_{\mu3}^{2}U_{e2}^{2}U_{\tau2}U_{\tau3}}{U_{\mu3}^{2}U_{e1}U_{\tau1}U_{\tau2}U_{e3}-U_{\mu1}^{2}U_{e3}^{2}U_{\tau2}U_{\tau3}}\right)\right)$,\\
  & \\
      & $\sigma=-\frac{1}{2}\left(\beta_{23}-\pi+Arg\left(\frac{U_{\mu3}^{2}U_{e2}^{2}U_{\tau2}U_{\tau3}-U_{\mu2}^{2}U_{e2}U_{e3}U_{\tau3}^{2}}{U_{\mu2}^{2}U_{e1}U_{\tau1}U_{\tau2}U_{e3}-U_{\mu1}^{2}U_{e2}U_{\tau2}^{2}U_{e3}}\right)\right)-\delta$.\\
     & \\
\hline
  $ B_{2}$   &$\rho=-\frac{1}{2}\left(\gamma_{23}-\pi+Arg\left(\frac{U_{\tau2}^{2}U_{e3}^{2}U_{\mu3}U_{\mu2}-U_{\tau3}^{2}U_{e2}U_{\mu2}^{2}U_{e3}}{U_{\tau3}^{2}U_{e1}U_{\mu1}U_{e2}U_{\mu3}-U_{\tau1}^{2}U_{e3}U_{\mu3}^{2}U_{e2}}\right)\right)$,\\
   &\\
       &$\sigma=-\frac{1}{2}\left(\gamma_{23}-\pi+Arg\left(\frac{U_{\tau3}^{2}U_{e2}U_{\mu2}^{2}U_{e3}-U_{\tau2}^{2}U_{e3}^{2}U_{\mu2}U_{\mu3}}{U_{\tau2}^{2}U_{e1}U_{\mu1}U_{e2}U_{\mu3}-U_{\tau1}^{2}U_{e2}^{2}U_{\mu2}U_{\mu3}}\right)\right)-\delta$.\\
       &\\
       
 \hline
  $ B_{3}$  &$\rho=-\frac{1}{2}\left(\gamma_{31}-\pi+Arg\left(\frac{U_{\mu2}^{2}U_{e3}U_{\mu3}^{2}U_{e1}-U_{e2}U_{\mu2}U_{\mu3}^{3}U_{e1}}{U_{\mu3}^{2}U_{e1}U_{\mu1}^{2}U_{e3}-U_{e3}^{2}U_{\mu3}U_{\mu1}^{3}}\right)\right)$,\\
   &\\
       &$\sigma=-\frac{1}{2}\left(\gamma_{31}-\pi+Arg\left(\frac{U_{\mu3}^{2}U_{e2}U_{\mu2}U_{e1}-U_{\mu2}^{2}U_{e3}U_{\mu3}^{2}U_{e1}}{U_{\mu2}^{2}U_{e1}U_{\mu1}U_{e3}-U_{e2}U_{\mu2}U_{\mu1}^{3}U_{e3}}\right)\right)-\delta$.\\
       &\\
\hline
  $ B_{4}$   & $\rho=-\frac{1}{2}\left(\beta_{31}-\pi+Arg\left(\frac{U_{\tau2}^{2}U_{e3}^{2}U_{\tau3}U_{\tau1}-U_{e2}U_{\tau2}U_{\tau3}^{2}U_{e3}U_{\tau1}}{U_{\tau3}^{3}U_{e1}^{2}U_{\tau1}U_{\tau3}-U_{e3}U_{\tau3}^{2}U_{\tau1}^{2}U_{e1}}\right)\right)$,   \\
   &\\
  &$\sigma=\frac{1}{2}\left(\beta_{31}-\pi+Arg\left(\frac{U_{\tau3}^{2}U_{e2}U_{\tau2}U_{\tau1}U_{e3}-U_{\tau2}^{2}U_{\tau1}U_{e3}^{2}U_{\tau3}}{U_{\tau2}^{2}U_{e1}^{2}U_{\tau1}U_{\tau3}-U_{e2}U_{\tau2}U_{\tau1}^{2}U_{\tau3}U_{e1}}\right)\right)-\delta$. \\
  &\\
  \hline
   $C$ &$\rho=\frac{1}{2}\left(\alpha_{31}-\pi+Arg\left(\frac{U_{\tau2}^{2}U_{\mu3}^{3}U_{\tau3}^{3}-U_{\mu2}^{2}U_{\tau3}^{5}U_{\mu3}}{U_{\mu1}^{3}U_{\tau3}^{2}U_{\tau1}^{3}-U_{\mu3}^{2}U_{\tau1}^{5}U_{\mu1}}\right)\right)$,\\
   &\\
   &$\sigma=\frac{1}{2}\left(\alpha_{31}-\pi+Arg\left(\frac{U_{\tau3}^{3}U_{\mu1}U_{\tau2}^2-U_{\mu3}^{2}U_{\tau2}^{2}U_{\mu1}U_{\tau3}}{U_{\mu1}^{2}U_{\tau2}^{2}U_{\mu3}U_{\tau1}-U_{\mu2}^{2}U_{\tau1}^{3}U_{\mu3}}\right)\right)-\delta$.\\
   &\\
\end{tabular}
\end{ruledtabular}
\end{table*}

We have, also, constructed the Majorana triangles for each type of two-texture zero neutrino mass matrices (Fig. \ref{fg2}, Fig. \ref{fg3} and Fig. \ref{fg4}) with best-fit values of neutrino mixing parameters shown in Table \ref{tab2}. 

\section{Status of Neutrinoless Double Beta Decay($0\nu\beta\beta$) in Two-Texture Zero Neutrino Mass Model}

The sensitivities of current and future experiments to Lepton Number Violating(LNV) processes  such as, $K^{+}\rightarrow \mu^{+}\mu^{+}\pi^{-}$ decay \cite{ndb1,ndb2,ndb3}, the nuclear muon to positron\cite{ndb4,ndb4a}, tri-muonium production in neutrino muon scattering\cite{ndb5}, the process $e^{+}p\rightarrow \bar{\nu}l_{1}^{+}l_{2}^{+}X$\cite{ndb6} and $0\nu\beta\beta$ decay\cite{ndbd,ndbda,ndbdb}, have been extensively studied in the literature and the existence of which will demonstrate the Majorana nature of the neutrinos. The sensitivities of the current experiments
searching for these processes are much less in comparison to experimental sensitivity to $0\nu\beta\beta$ decay process. $0\nu\beta\beta$ decay is the most promising probe of observing lepton number violation and will shed light on mechanism of neutrino mass generation. The effective Majorana mass, $\left<m\right>_{ee}$, may vanish for normal hierarchical neutrino masses, however, for inverted hierarchical neutrino masses there exist a lower bound on $\left<m\right>_{ee}$ providing bright prospects for observation of $0\nu\beta\beta$ decay. Although, it will be difficult to measure Majorana phases in these experiments but a correlation between them can, always, be obtained. One can uniquely determine both Majorana phases for the case of vanishing $\left<m\right>_{ee}$\cite{1608.01618}. The decay width of the $0\nu\beta\beta$ is proportional to the effective Majorana mass, $\left<m\right>_{ee}$, given by\cite{ndb7}
\begin{eqnarray}\label{mee1}
\nonumber
 \left<m\right>_{ee}&\equiv&\sum_{i=1}^{3} V_{ei}^{2}m_{i},\\ 
  &=&m_1\left| U_{e1}^2\right|+m_2 \left|U_{e2}^2\right| e^{2i\rho}+m_3\left|U_{e3}^2\right|e^{2i\sigma}.
\end{eqnarray}

Using the allowed parameter space obtained for two-texture zero neutrino mass matrices shown in Table \ref{tab2}, we have obtained the $1\sigma$  range of effective Majorana mass, $\left|\left<m\right>_{ee}\right|$, for each type of two-texture zero neutrino mass model which has been tabulated in Table \ref{tab5}. For class $A$ type neutrino mass models, the effective Majorana mass, $\left<m\right>_{ee}$ is vanishing thus observation of $0\nu\beta\beta$ decay will rule out these class of neutrino mass models. The sensitivities of the $0\nu\beta\beta$ experiments like KamLAND-ZEN \cite{kam_zen1,kam_zen2}, GERDA \cite{gerda}, CUORE \cite{cuore1,cuore2}, NEXT \cite{next1,next2} and EXO-200 \cite{exo1,exo2} have set strong bounds on effective Majorana mass, $\left|\left<m\right>_{ee}\right|$, tabulated in Table \ref{tab6}. Recently, the strongest constraint has been obtained by KamLAND-ZEN \cite{kam_zen2}. The predictions for effective Majorana mass, $\left|\left<m\right>_{ee}\right|$, for class $B$ and $C$ are found to be well within the sensitivity reach of current and future $0\nu\beta\beta$ decay experiments(Table \ref{tab6}). However, non-observation of $0\nu\beta\beta$ decay will favour class $A$ and class $B$ (with NH) type neutrino mass models. 

\begin{table}[h]
 \caption{\label{tab5} $\left|\left<m\right>_{ee}\right|$(eV) for each type of two-texture zero neutrino mass matrices.}
 \begin{ruledtabular}
  \centering
 \begin{tabular}{c  l  l}

  {Texture} & NH(bfp $\pm1\sigma$)&  IH(bfp $\pm1\sigma$) \\
  \hline
    $A_{1}$ & $\left|\left<m\right>_{ee}\right|=0$& -\\ 
             \hline
   $ A_{2}$ & $\left|\left<m\right>_{ee}\right|=0$& -\\ 
    \hline
    $B_{1}$ & $\left|\left<m\right>_{ee}\right|=0.059\pm 0.016$&$\left|\left<m\right>_{ee}\right|=0.181\pm0.066$\\
                 \hline 
    $B_{2}$ & $\left|\left<m\right>_{ee}\right|=0.173\pm0.088$&$\left|\left<m\right>_{ee}\right|=0.080\pm0.014$\\
                      \hline
   $ B_{3}$ &$\left|\left<m\right>_{ee}\right|=0.061\pm0.016$ &$\left|\left<m\right>_{ee}\right|=0.033\pm 0.015$\\
   \hline
   $ B_{4}$ & $\left|\left<m\right>_{ee}\right|=0.180\pm0.113$ & $\left|\left<m\right>_{ee}\right|=0.084\pm0.016$\\
    \hline
   $ C$ &  - & $\left|\left<m\right>_{ee}\right|=0.065\pm0.043$\\
  
  \end{tabular}
  \end{ruledtabular}
\end{table}

\begin{table}[h]
 \caption{\label{tab6} Sensitivities to effective Majorana mass, $\left|\left<m\right>_{ee}\right|$, of various $0\nu\beta\beta$ decay experiments\cite{expts}.}
 \begin{ruledtabular}
  \centering
 \begin{tabular}{ l  l}
  {Experiment} & $\left|\left<m\right>_{ee}\right|$(eV) \\
  \hline
    EXO-200(4 yr) & 0.075-0.2\\ 
             \hline
    nEXO(5 yr) & 0.012-0.029\\ 
    \hline
    nEXO(5 yr + 5 yr w / Ba tagging) & 0.005-0.011\\
                 \hline 
    KamLAND-ZEN(300 kg, 3 yr) & 0.045-0.11\\
                      \hline
   GERDA phase II & 0.09-0.29\\
   \hline
   CUORE(5 yr) & 0.051-0.133\\
    \hline
   SNO+ &  0.07-0.14\\
    \hline
   SuperNEMO &  0.05-0.15\\
    \hline
   NEXT &  0.03-0.1\\
    \hline
   MAJORANA Demonstrator &  0.06-0.17\\

  \end{tabular}
  \end{ruledtabular}
\end{table}
\section{Results and Discussion}
In class $A$, $\cos\delta$ should be positive(negative) for $A_1$($A_2$) because $\frac{m_1}{m_2}<1$. So, $\delta$ can be in I or IV quadrant for $A_{1}$ and in II and III quadrant for $A_2$ type mass matrix. These theoretical predictions are consistent with the obtained best-fit values of $\delta$ shown in Table \ref{tab2}.

The class $B$ yield a quasi-degenerate spectrum of neutrino masses. For mass ratio $\frac{m_{1}}{m_{2}}<1$, $\cos\delta$ should be negative(positive) for $B_{1}$ and $B_{4}$($B_{2}$ and $B_{3}$). We find that $\delta=267.10^{o}, 266.91^{o}, 267.80^{o}, 268.50^{o}$ ($ 268.87^{o}, 269.00^{o}, \\269.47^{o}, 266.82^{o}$) for $B_1, B_2, B_3$ and $B_4$ with NH(IH), respectively. The important point here is to note that the best-fit values of $\delta$ comes out to be nearly equal to $\frac{3\pi}{2}$ for  class B, which is in accordance with the experimental value of $\delta$ predicted by the combined analysis of T2K and Daya Bay experiments\cite{capozzi}. 

\begin{figure*}
\vspace{8cm}
\centering
 \includegraphics[scale=0.625]{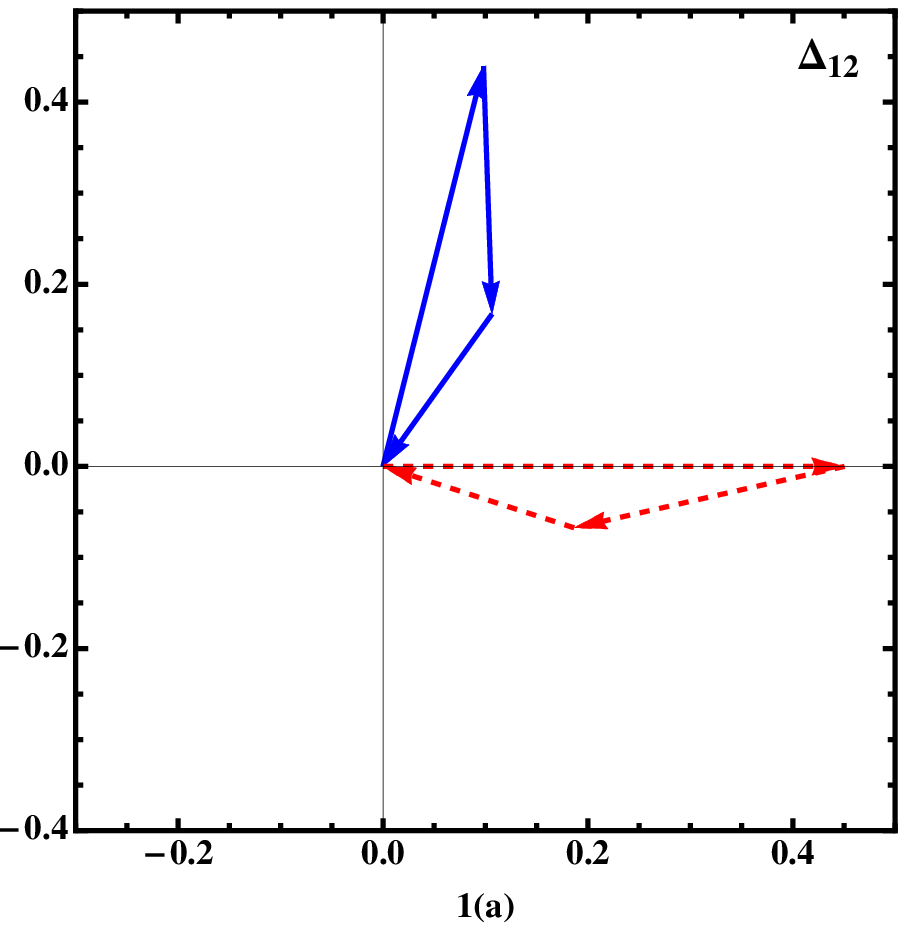}\hspace{2cm}
  \includegraphics[scale=0.60]{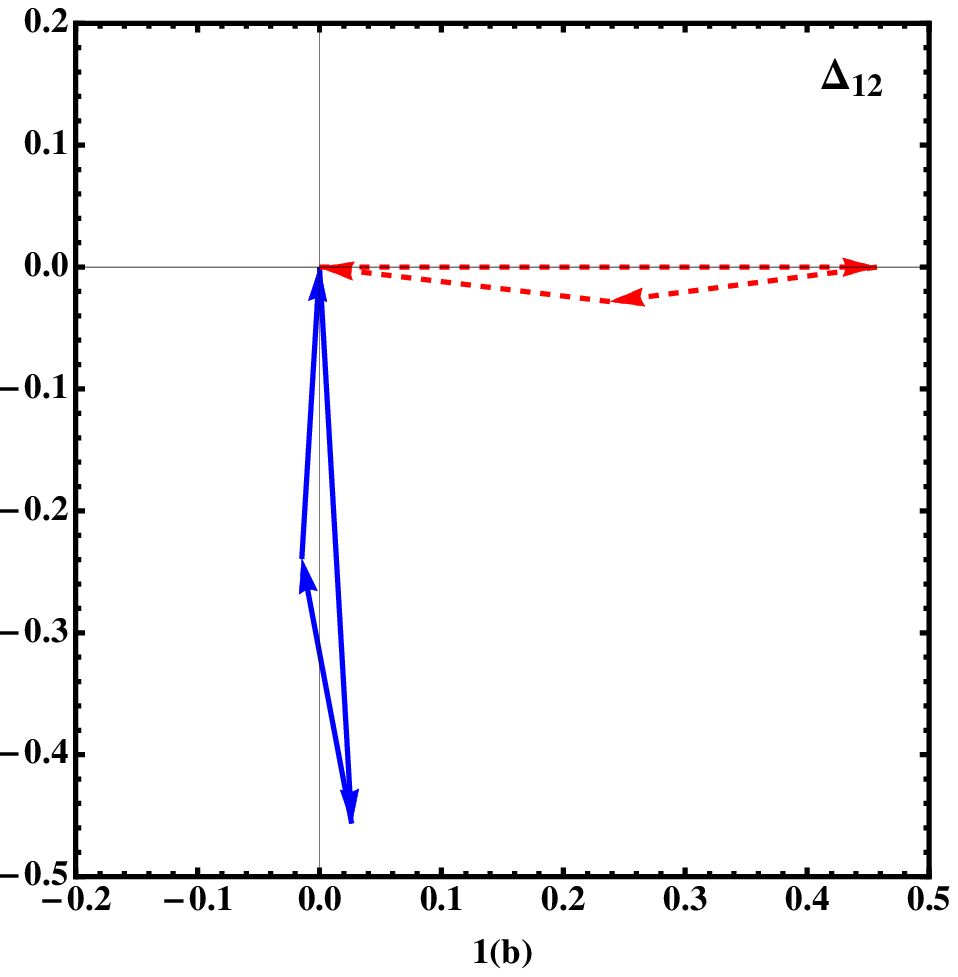}
 \caption{\label{fg2}Majorana triangles for class $A$ neutrino mass matrices with normal hierarchical(NH) neutrino masses. MT for type $A_{1}(A_{2})$ with NH in left(right).}
\end{figure*}

\begin{figure*}
\centering
 \includegraphics[scale=0.8]{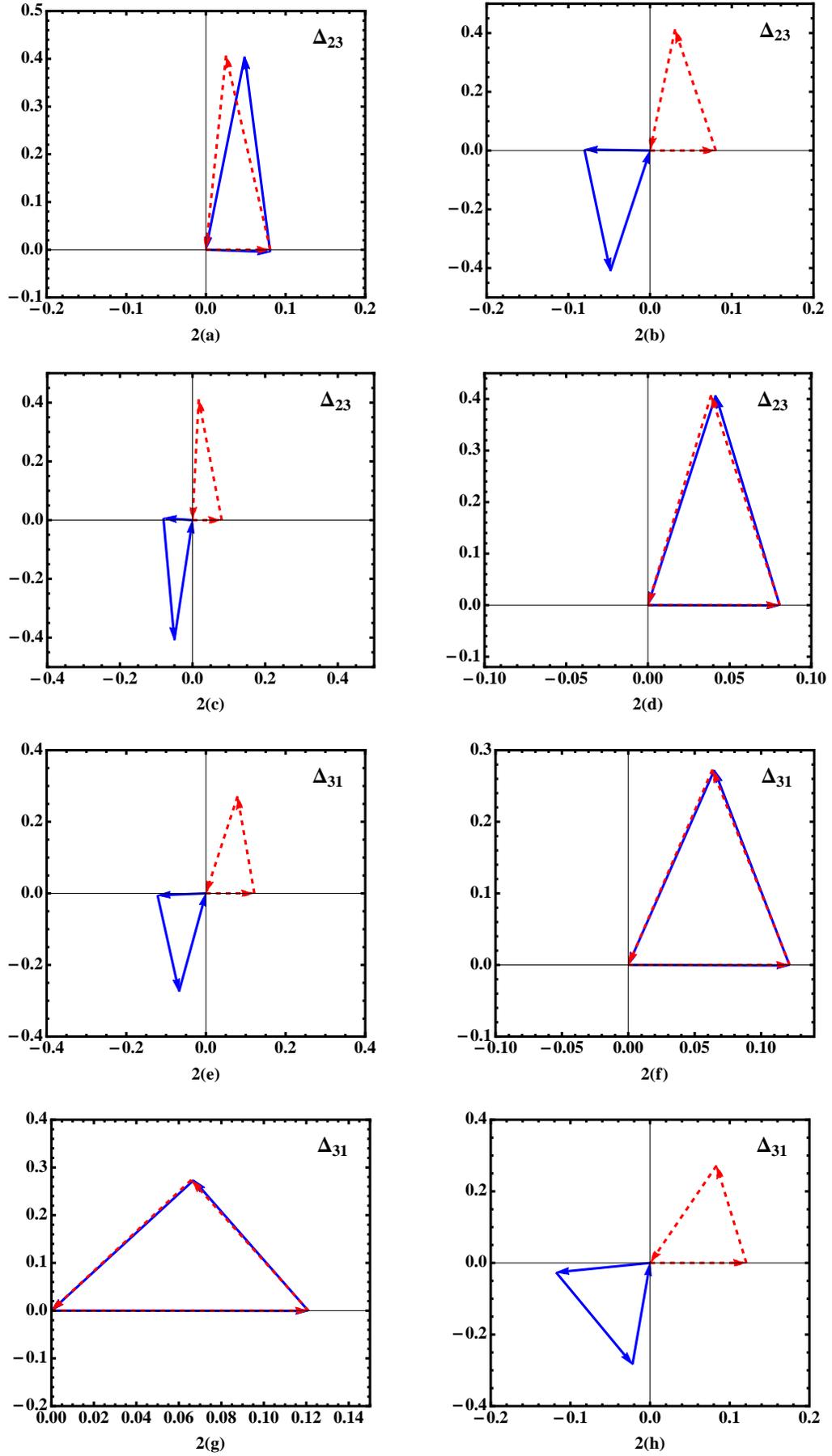}
  \caption{\label{fg3}Majorana triangles for class $B$ neutrino mass matrices with normal and inverted hierarchical(IH) neutrino masses. MT for $B_{1},B_{2},B_{3}$ and $B_{4}$ with NH(left) and IH(right).}
\end{figure*}

\begin{figure*}
 \includegraphics[scale=0.8]{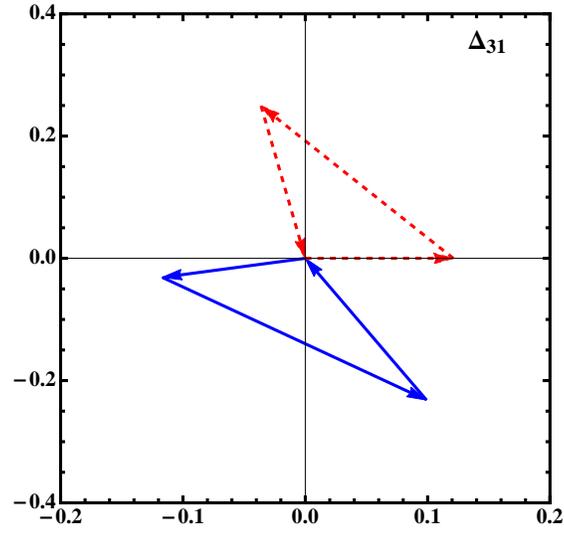}
 \caption{\label{fg4}Majorana triangles for class $C$ neutrino mass matrices with inverted hierarchical neutrino masses.}
\end{figure*}

In class $C$, for $\frac{m_1}{m_2}<1$, the factor $\tan2\theta_{23}\cos\delta$ should be positive and $\delta$ must lie either in I or IV quadrant. This prediction is found to be consistent with best-fit value of $\delta=292.80^{o}$(Table \ref{tab2}).
In Table \ref{tab3}, we have obtained the $CP$ invariants corresponding to Dirac and Majorana type $CP$ violation for class $A,B$ and $C$ neutrino mass matrices. The existence of non-zero $CP$ violation is a generic prediction of two-texture zero neutrino mass models.  
In Table \ref{tab4}, we have, also, obtained the expressions for Majorana phases in terms of the interior angle of Majorana unitarity triangle for class $A$, $B$ and $C$. We observe that the Majorana phases depend only on one independent geometric parameter of MT for each class of two-texture zero neutrino mass model. The orientation of these triangles depend on the Majorana phases as,
\begin{eqnarray}\label{majoranatriangles}
\nonumber
\Delta_{12}&&\equiv e^{i\rho}\left(U_{e1}U_{e2}^{*}+U_{\mu1}U_{\mu2}^{*}+U_{\tau1}U_{\tau2}^{*}\right),\\
\nonumber
\Delta_{23}&&\equiv e^{i(\delta+\sigma-\rho)}\left(U_{e2}U_{e3}^{*}+U_{\mu2}U_{\mu3}^{*}+U_{\tau2}U_{\tau3}^{*}\right),\\
\Delta_{31}&&\equiv e^{i(\delta+\sigma)}\left(U_{e3}U_{e1}^{*}+U_{\mu3}U_{\mu1}^{*}+U_{\tau3}U_{\tau1}^{*}\right).
\end{eqnarray}
Using Eq.(\ref{majoranatriangles}) alongwith values of mixing parameters as shown in Table \ref{tab2}, the Majorana triangles have been constructed as shown with solid lines in Figs. (\ref{fg2}-\ref{fg4}). For reference, the dashed triangles in  Figs. (\ref{fg2}-\ref{fg4}) are obtained assuming $\rho$ and $\sigma$ equal to zero. In general, we find that the Majorana triangles provide an indubitable signal towards non-zero $CP$ violation in this class of models because none of the sides of the triangle is parallel to the axis and all the Majorana triangles having non-vanishing area.

\section{Conclusions}
In conclusion, we have investigated $CP$ violation in neutrino mass models with two-texture zeros. In particular, we have obtained possible connection between Majorana phases($\rho, \sigma$) and independent geometric parameters of Majorana triangle(MT). We find that Majorana phases depend on one interior angle of MT i.e. $\gamma_{12}$ in $A_1$, $\beta_{12}$ in $A_2$, $\beta_{23}$ in $B_1$, $\gamma_{23}$ in $B_2$, $\gamma_{31}$ in $B_3$, $\beta_{31}$ in $B_4$ and $\alpha_{31}$ in class $C$ type neutrino mass matrix. This analysis is important in light of the future neutrino oscillation experiments focussing on measuring Dirac-type $CP$ violation phase $\delta$. We  have, also, obtained the best-fit and $\pm1\sigma$ values of neutrino oscillation parameters and $CP$ rephasing invariants for class $A, B$ and $C$ neutrino mass matrices. We find that for class $B$, the best-fit value lies very close to $\delta\approx\frac{3\pi}{2}$ which is in accordance with T2K and Daya Bay experiments\cite{capozzi}. The non-zero area and non-trivial orientation of Majorana triangles shows that two-texture zero neutrino mass matrices are necessarily $CP$ violating. 

The current and future $0\nu\beta\beta$ decay experiments can shed light on two-texture zero neutrino mass models. As long as fixing of parameters(absolute neutrino mass and Majorana phases) is concerned, the non-observation of $0\nu\beta\beta$ decay is more predictive than the situation in which $0\nu\beta\beta$ decay is observed. The predictions for effective Majorana mass, $\left|\left<m\right>_{ee}\right|$, for class $B$ and $C$ are found to be well within the sensitivity reach of current and future $0\nu\beta\beta$ decay experiments(Table \ref{tab6}). The non-observation of $0\nu\beta\beta$ decay will favour class $A$ and class $B$ (with NH) type two-texture zero neutrino mass matrices while class $C$ will be ruled out. 
 

\begin{acknowledgments}
S. V. acknowledges the  financial support provided by University Grants Commission (UGC)-Basic Science
Research(BSR), Government of India vide Grant No. F.20-2(03)/2013(BSR). S. B. acknowledges the financial 
support provided by the Central University of Himachal Pradesh. 
\end{acknowledgments}
\appendix*
\section{Majorana phases in terms of interior angles of the leptonic unitarity triangle}
\begin{figure*}[h]
 	\centering
 	\includegraphics[scale=0.4]{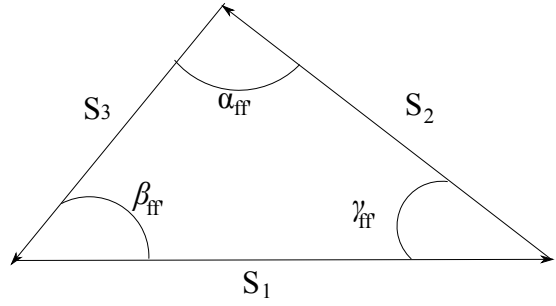}
 	\caption{\label{fg1}Majorana Unitarity Triangle in complex plane.}
 \end{figure*}

 We have obtained relations of Majorana phases $\rho$ and $\sigma$ in terms of geometric angles of Majorana triangle for each type of two-texture zero neutrino mass matrices(Table \ref{tab4}). The method for type $A_{1}$ has been elaborated here. We choose triangle as shown in Fig.\ref{fg1} with $f=1$, $f'=2$ and angles of triangle as shown in Eq.(\ref{angles}). 

\begin{widetext}
 \begin{eqnarray*}
\nonumber
\rho &&= -\frac{1}{2}\biggl(Arg\left(\frac{U_{e2}U_{\mu2}U_{e3}^{2}-U_{e2}^{2}U_{e3}U_{\mu3}}{U_{e3}U_{\mu3}U_{e1}^{2}-U_{e3}^{2}U_{e1}U_{\mu1}}\right)\biggr)\\
\nonumber
     &&= -\frac{1}{2}\biggl(Arg\biggl(\left(\frac{U_{e2}U_{\mu2}}{U_{e3}U_{\mu3}}\right)\left(\frac{U_{e3}U_{\mu3}}{U_{e1}U_{\mu1}}\right)
    \left(\frac{U_{e3}U_{\mu2}-U_{e2}U_{\mu3}}{U_{e1}U_{\mu3}-U_{e3}U_{\mu1}}\right) \left(\frac{U_{\mu1}U_{\mu3}}{U_{\mu3}U_{\mu2}}\right)\biggr)\biggr)\\
\nonumber
     &&=-\frac{1}{2}\left(Arg\left(\frac{U_{e2}U_{\mu2}}{U_{e1}U_{\mu1}}\right)+Arg\left(\frac{U_{e3}U_{\mu2}U_{\mu1}-U_{e2}U_{\mu3}U_{\mu1}}{U_{\mu2}U_{e1}U_{\mu3}-U_{\mu2}U_{e3}U_{\mu1}}\right)\right)\\
     \nonumber
     &&=-\frac{1}{2}\left(Arg\left(\frac{U_{e2}}{U_{\mu1}}\frac{U_{\mu2}}{U_{e1}}\right)+Arg\left(\frac{U_{e3}U_{\mu2}U_{\mu1}-U_{e2}U_{\mu3}U_{\mu1}}{U_{\mu2}U_{e1}U_{\mu3}-U_{\mu2}U_{e3}U_{\mu1}}\right)\right)\\
     \nonumber
     &&=-\frac{1}{2}\biggl(Arg\left(\frac{U_{e2}}{U_{\mu1}}\frac{U_{e1}}{U_{\mu2}}\right)+Arg\left(\frac{U_{e3}U_{\mu2}U_{\mu1}-U_{e2}U_{\mu3}U_{\mu1}}{U_{\mu2}U_{e1}U_{\mu3}-U_{\mu2}U_{e3}U_{\mu1}}\right)+2 Arg\left(\frac{U_{\mu2}}{U_{e1}}\right)\biggr)\\
     \nonumber
     &&=-\frac{1}{2}\biggl(Arg\biggl(\frac{U_{e1}}{U_{\mu1}}\frac{U_{e2}}{U_{\mu2}}+Arg\left(\frac{U_{e3}U_{\mu2}U_{\mu1}-U_{e2}U_{\mu3}U_{\mu1}}{U_{\mu2}U_{e1}U_{\mu3}-U_{\mu2}U_{e3}U_{\mu1}}\right)+2 Arg\left(\frac{U_{\mu2}}{U_{e1}}\right)\biggr)\biggr)\\
     \nonumber
     &&=-\frac{1}{2}\biggl(Arg\left(-\frac{U_{e1}}{U_{\mu1}}\frac{U_{e2}^{*}}{U_{\mu2}^{*}}\right)-\pi+2 Arg\left(\frac{U_{e2}}{U_{\mu2}}\right)+2Arg\left(\frac{U_{\mu2}}{U_{e1}}\right)+Arg\left(\frac{U_{e3}U_{\mu2}U_{\mu1}-U_{e2}U_{\mu3}U_{\mu1}}{U_{\mu2}U_{e1}U_{\mu3}-U_{\mu2}U_{e3}U_{\mu1}}\right)\biggr)\\
       \nonumber
     &&=-\frac{1}{2}\biggl(Arg\left(-\frac{U_{e1}U_{e2}^{*}}{U_{\mu1}U_{\mu2}^{*}}\right)-\pi+2 Arg\left(\frac{U_{e2}U_{\mu2}}{U_{e1}U_{\mu2}}\right)+Arg\left(\frac{U_{e3}U_{\mu2}U_{\mu1}-U_{e2}U_{\mu3}U_{\mu1}}{U_{\mu2}U_{e1}U_{\mu3}-U_{\mu2}U_{e3}U_{\mu1}}\right)\biggr)\\
       \nonumber
     &&=-\frac{1}{2}\biggl(Arg\left(-\frac{U_{e1}U_{e2}^{*}}{U_{\mu1}U_{\mu2}^{*}}\right)-\pi+Arg\biggl(\left(\frac{U_{e3}U_{\mu2}U_{\mu1}-U_{e2}U_{\mu3}U_{\mu1}}{U_{\mu2}U_{e1}U_{\mu3}-U_{\mu2}U_{e3}U_{\mu1}}\right)\left(\frac{U_{e2}^{2}}{U_{e1}^{2}}\right)\biggr)\biggr)\\
     \nonumber
     &&=-\frac{1}{2}\biggl(\gamma_{12}-\pi+Arg\left(\frac{U_{e3}U_{\mu2}U_{\mu1}U_{e2}^{2}-U_{e2}^{3}U_{\mu3}U_{\mu1}}{U_{\mu2}U_{e1}^{3}U_{\mu3}-U_{\mu2}U_{e3}U_{\mu1}U_{e1}^{2}}\right)\biggr).
   \end{eqnarray*}
   \end{widetext}
where we have used the identities, $Arg\left(\frac{z_{1}}{z_{2}}\right)=Arg(z_{1})-Arg(z_{2}), Arg(z_{1}z_{2})=Arg(z_{1})+Arg(z_{2}), Arg\left(\frac{1}{z}\right)=Arg(z^{*})=-Arg(z), Arg(-z)=Arg(z)+\pi, Arg(z^{n})=n Arg(z)$. Note that all these equalities hold modulo $2n\pi(n\in\mathbb{Z})$.  
 
 \begin{widetext}  
 \begin{eqnarray*}
\nonumber
\sigma &&= -\frac{1}{2}\left(Arg\left(\frac{U_{e3}U_{\mu3}U_{e2}^{2}-U_{e3}^{2}U_{e2}U_{\mu2}}{U_{e2}U_{\mu2}U_{e1}^{2}-U_{e2}^{2}U_{e1}U_{\mu1}}\right)\right)-\delta\\
\nonumber
       && = -\frac{1}{2}\left(Arg\left(\frac{U_{e2}U_{\mu2}}{U_{e1}U_{\mu1}}\right)+Arg\left(\frac{\frac{U_{e2}^{2}U_{e3}U_{\mu3}}{U_{e2}U_{\mu2}}-U_{e3}^{2}}{\frac{U_{e1}^{2}U_{e2}U_{\mu2}}{U_{e1}U_{\mu1}}-U_{e2}^{2}}\right)\right)-\delta\\
     \nonumber
     &&=-\frac{1}{2}\left(Arg\left(\frac{U_{e2}}{U_{\mu1}}\frac{U_{\mu2}}{U_{e1}}\right)+Arg\left(\frac{\frac{U_{e2}U_{e3}U_{\mu3}}{U_{\mu2}}-U_{e3}^{2}}{\frac{U_{e1}U_{e2}U_{\mu2}}{U_{\mu1}}-U_{e2}^{2}}\right)\right)-\delta\\
     \nonumber
     &&=-\frac{1}{2}\left(Arg\left(\frac{U_{e2}}{U_{\mu1}}\frac{U_{e1}}{U_{\mu2}}\right)+2 Arg\left(\frac{U_{\mu2}}{U_{e1}}\right)+Arg\left(\frac{\frac{U_{e2}U_{e3}U_{\mu3}}{U_{\mu2}}-U_{e3}^{2}}{\frac{U_{e1}U_{e2}U_{\mu2}}{U_{\mu1}}-U_{e2}^{2}}\right)\right)-\delta\\
     \nonumber
     &&=-\frac{1}{2}\left(Arg\left(\frac{U_{e1}}{U_{\mu1}}\frac{U_{e2}}{U_{\mu2}}\right)+2 Arg\left(\frac{U_{\mu2}}{U_{e1}}\right)+Arg\left(\frac{\frac{U_{e2}U_{e3}U_{\mu3}}{U_{\mu2}}-U_{e3}^{2}}{\frac{U_{e1}U_{e2}U_{\mu2}}{U_{\mu1}}-U_{e2}^{2}}\right)\right)-\delta\\
     \nonumber
       &&=-\frac{1}{2}\left(Arg\left(-\frac{U_{e1}}{U_{\mu1}}\frac{U_{e2}^{*}}{U_{\mu2}^{*}}\right)-\pi+2 Arg\left(\frac{U_{e2}}{U_{\mu2}}\right)+2Arg\left(\frac{U_{\mu2}}{U_{e1}}\right)+Arg\left(\frac{\frac{U_{e2}U_{e3}U_{\mu3}}{U_{\mu2}}-U_{e3}^{2}}{\frac{U_{e1}U_{e2}U_{\mu2}}{U_{\mu1}}-U_{e2}^{2}}\right)\right)-\delta\\
       \nonumber
       &&=-\frac{1}{2}\left(Arg\left(-\frac{U_{e1}U_{e2}^{*}}{U_{\mu1}U_{\mu2}^{*}}\right)-\pi+2 Arg\left(\frac{U_{e2}U_{\mu2}}{U_{e1}U_{\mu2}}\right)+Arg\left(\frac{\frac{U_{e2}U_{e3}U_{\mu3}}{U_{\mu2}}-U_{e3}^{2}}{\frac{U_{e1}U_{e2}U_{\mu2}}{U_{\mu1}}-U_{e2}^{2}}\right)\right)-\delta\\
       \nonumber
       &&=-\frac{1}{2}\left(Arg\left(-\frac{U_{e1}U_{e2}^{*}}{U_{\mu1}U_{\mu2}^{*}}\right)-\pi+Arg\left(\left(\frac{\frac{U_{e2}U_{e3}U_{\mu3}}{U_{\mu2}}-U_{e3}^{2}}{\frac{U_{e1}U_{e2}U_{\mu2}}{U_{\mu1}}-U_{e2}^{2}}\right)\left(\frac{U_{e2}^{2}}{U_{e1}^{2}}\right)\right)\right)-\delta\\
       \nonumber
       &&=-\frac{1}{2}\left(\gamma_{12}-\pi+Arg\left(\frac{\frac{U_{e2}^{3}U_{e3}U_{\mu3}}{U_{\mu2}}-U_{e3}^{2}U_{e2}^{2}}{\frac{U_{e1}^{3}U_{e2}U_{\mu2}}{U_{\mu1}}-U_{e2}^{2}U_{e1}^{2}}\right)\right)-\delta\\
       \nonumber
       &&=-\frac{1}{2}\left(\gamma_{12}-\pi+Arg\left(\left(\frac{U_{e2}^{3}U_{e3}U_{\mu3}-U_{e3}^{2}U_{e2}^{2}U_{\mu2}}{U_{e1}^{3}U_{e2}U_{\mu2}-U_{e2}^{2}U_{e1}^{2}U_{\mu1}}\right)\left(\frac{U_{\mu1}}{U_{\mu2}}\right)\right)\right)-\delta\\
       &&=-\frac{1}{2}\left(\gamma_{12}-\pi+Arg\left(\frac{U_{e2}^{3}U_{e3}U_{\mu3}U_{\mu1}-U_{e3}^{2}U_{e2}^{2}U_{\mu2}U_{\mu1}}{U_{e1}^{3}U_{e2}U_{\mu2}^{2}-U_{e2}^{2}U_{e1}^{2}U_{\mu1}U_{\mu2}}\right)\right)-\delta.
\end{eqnarray*}
\end{widetext}
In similar way, we can obtain expressions for other type of 
textures $A_{2}, B_{1}, B_{2}, B_{3}, B_{4}$ and $C$.

\end{document}